\documentstyle[aps,twocolumn]{revtex}

\begin{document}
\title{Chaos, thermodynamics and quantum mechanics: 
an application to celestial dynamics}
\author{David Vitali~$^1$ and Paolo Grigolini~$^{2,3,4}$}
\address {$^*$~Dipartimento di Matematica e Fisica, Universit\`a di
Camerino, via Madonna delle Carceri I-62032 Camerino \\
and Istituto Nazionale di Fisica della Materia, Camerino, Italy\\
$^1$~Dipartimento di Fisica, Universit\`{a} di Pisa,
Piazza Torricelli 2, 56126 Pisa, Italy\\
$^2$~Istituto di Biofisica del Consiglio Nazionale delle
Ricerche,
via San Lorenzo 28, 56127 Pisa, Italy \\
$^3$~Department of Physics, University of North Texas,
P.O. Box 5638, Denton, Texas 76203}
\maketitle

\begin{abstract}
We address the issue of the quantum-classical correspondence in chaotic 
systems using, as recently done by Zurek [e-print quant-ph/9802054], the 
solar system as a whole as a case study: this author shows that the 
classicality of the planetary motion is ensured by the environment-induced 
decoherence.  We show that equivalent results are provided by the theories 
of spontaneous fluctuations and that these latter theories, in some cases, 
result in a still faster process of decoherence.  We show that, as an 
additional benefit, the assumption of spontaneous fluctuation makes it 
possible to genuinely derive thermodynamics from mechanics, namely, without 
implicitly assuming thermodynamics.
\end{abstract}

\section{Introduction}

The problem of how the classical macroscopic world emerges from the quantum 
substrate is the main problem of the interpretation of quantum mechanics 
and it is still the subject of an intense debate \cite{ZEH98}.  A well 
known conceptual difficulty stems from the quantum states which are linear 
superpositions of macroscopically distinguishable properties.  Although 
these states have never been observed, they are, according to quantum 
mechanics, a legitimate physical condition, thereby resulting in the well 
known paradox of the Schr\"odinger cat \cite{SCH35} and, more in general, of 
the measurement processes \cite{WZ83}.  

Another controversial, and more 
subtle, problem is as to whether or not the correspondence principle, 
namely, the alleged physical property that makes the quantum mechanical 
predictions on macroscopic systems identical to those of classical 
mechanics, applies to the crucial case of classically chaotic systems.  
There exists a general agreement on the following fact: a quantum system 
with a classical-like initial condition, i.e.  a wave packet localized and 
smooth over scales much larger than $\hbar$, will depart, sooner or later, 
from this classical condition as a natural effect of time evolution.  The 
object of debate is given by the estimate of this time.  According to some 
authors (see \cite{CC95} and references therein), relevant discrepancies 
between classical and quantum motion occur only after a ``relaxation time'' 
$t_{R}$ proportional to some (system-dependent) power of the ratio 
$A/\hbar$, 
where $A$ is some characteristic action.  For macroscopic systems this ratio 
is always so large as to make the correspondence principle essentially 
safe.  According to Zurek and Paz \cite{ZP94,ZP96} and Zurek \cite{Z98} 
instead, the quantum systems that would be classically chaotic, are 
expected to depart from the classical-like initial condition at a time 
depending logarithmically on the ratio $A/\hbar$, $t_{Q} \simeq 
\lambda^{-1} \ln(A/\hbar)$ ($\lambda$ is 
the relevant Lyapunov exponent) and consequently much sooner. This 
logarithmic time is denoted by the subscript $Q$ to stress its meaning of 
transition from classical to quantum mechanics.  This time can be 
uncomfortably short even for macroscopic systems, so short as to result in 
a new kind of quantum paradox \cite{G93}: deterministic chaos would establish 
quantum correlations among macroscopically distinct regions of the 
classical phase space of these systems.  

This new paradox would break the 
correspondence principle and Zurek and Paz \cite{ZP94,ZP96} and Zurek 
\cite{Z98} state that in this case an efficient way of settling the 
paradox is offered by the theory of environmental decoherence 
\cite{ZEH98,Z81}.  The tenet of this theory is based on the observation 
that it is impossible to isolate completely a macroscopic system from the 
environment and that this unescapable interaction results in fluctuations, 
which, in turn, result in the rapid decay of the quantum coherence among 
macroscopically distinct components of the same quantum state.  Decoherence 
quickly transforms a linear superposition of macroscopically 
distinguishable properties, for instance positions, into the corresponding 
classical statistical mixture; it represents therefore the {\em 
practical} 
solution of the Schr\"odinger cat paradox, even if from a conceptual and 
interpretational point of view, many problems are still open (see for 
example \cite{PT93,BGM95}).  

Zurek and Paz support this view with 
suggestive examples borrowed from celestial dynamics \cite{ZP96,Z98}.  
There is now in fact well-established numerical evidence of chaotic motion 
of celestial bodies, such as the chaotic tumbling of Hyperion, one of the 
moons of Saturn \cite{WPM84}, the orbital motion of Chiron in the asteroid 
belt \cite{W85}, and the dynamics of the solar system as a whole 
\cite{L89,SW92}.  Quantum mechanics is judged to be more general than 
classical physics; furthermore the elementary constituents of these systems 
must obey quantum laws.  Thus, we are forced to consider these systems as 
being quantum, and we are led to estimate the corresponding logarithmic 
time.  The surprising conclusion is that this time turns out to be much 
shorter than the age of the systems themselves, since $t_{Q} \simeq 7 
\cdot 10^{8}$ years for 
the solar system and $t_{Q} \simeq 20$ years for Hyperion.  

According to quantum 
mechanics, as a consequence of this fact, the present state of Hyperion 
should be a coherent superposition of macroscopically distinct orientations 
of its major axis and, much more paradoxically, in the case of the planets 
of the solar system the wave function should have spread all over their 
orbits!  The solution of the paradox given by Zurek and Paz in 
\cite{ZP96,Z98} is based on decoherence: even the tiny interaction with the 
interplanetary gas is sufficient to achieve the destruction of quantum 
coherence over macroscopic distances so as to obtain a localized wavepacket 
moving according to Newton laws.  This takes place at a time much smaller 
than the logarithmic time $t_{Q}$ and any transition from classical to quantum 
mechanics is made impossible afterward.  

Several aspects of the proposal by 
Zurek and Paz to make the quantum-classical correspondence safe by means of 
the environmental fluctuations have been recently criticized 
\cite{CC95,FBA96}.  For this reason, in this letter we plan to derive first 
of all the same physical conclusions from an approach bypassing, in part, 
these criticisms.  We shall show that the time $t_{Q}$ is the correct estimate 
for the correspondence breakdown time if the condition of strong chaos case 
is ensured (i.e.  when the classical phase space is filled by the chaotic 
sea with no appreciable stable island) and we shall consequently provide 
new arguments in favor of fluctuations as a proper route to the 
correspondence principle.  However, and this is an other element of novelty 
of this letter, we shall consider also the possibility that these essential 
fluctuations might have a non-environmental origin.  This means that we 
shall consider some models of spontaneous wave-function reduction 
\cite{GRW,GPR,PD,GGR}.  These theories generalize quantum mechanics by 
adding stochastic and nonlinear terms to the Schr\"odinger equation.  We 
shall show that all these models make the solar system obey the 
quantum-classical correspondence, in the same way as the theory of 
environment-induced decoherence does.  In particular the 
gravitation-induced collapse theory, a model originally introduced 
by Di\'osi and Penrose \cite{PD}, and later improved by Ghirardi, Grassi and 
Rimini \cite{GGR}, turns out to be more efficient than the environmental 
process discussed by Zurek in \cite{Z98}, yielding in fact a much faster 
transition from the quantum to the classical domain.  This is of special 
interest because this is the first dynamical system where the decoherence 
induced by spontaneous fluctuations is shown to become more important than 
the environmental decoherence \cite{BGM95,TEG91,TVG95,BJK98}.  

To properly 
point out the significance of this aspect, it is worth mentioning one of 
the most frequent criticisms of the theories of spontaneous wave-functions 
collapses.  All these generalizations of quantum mechanics have been 
developed by assigning {\it ad hoc} values to the parameters establishing the 
strength of the corrections to ordinary quantum mechanics.  These choices 
serve the important purpose of making these theories pass all possible 
experimental tests (namely to coincide with quantum mechanics where this 
theory agrees with experimental results).  As a consequence, the intensity 
of the corrections is so weak as to make it difficult to prove them with 
experiments.  Furthermore, their key action is usually blurred by 
environmental decoherence, which is expected to be always predominant 
\cite{TEG91,TVG95,BJK98}.  We show that in the case of solar system it is 
not so.  

Of course, this conclusion does not rule out the theory of 
environmental decoherence, since it is not possible to design an experiment 
aiming at assessing whether the solar system is made classical by 
environmental or spontaneous fluctuations.  As argued in Section IV, we 
think that the choice of the theory of spontaneous fluctuations can be done 
only on the basis of theoretical arguments concerning the problem of 
unification of thermodynamics and mechanics as well as that of classical 
and quantum mechanics.  

The structure of this letter is as follows.  In 
Section II we briefly review the original work of Zurek and Paz \cite{ZP94} 
and some of the criticisms raised by this work \cite{CC95,FBA96}.  Section 
III illustrates more general heuristic arguments, which yield the same 
conclusions as those by Zurek and Paz in the case of strong chaos.  In this 
Section we also point out the essential role of coarse-graining processes 
for the foundation of thermodynamics.  Section IV shows that the theory of 
spontaneous fluctuations leads to results formally equivalent to those of 
Section II, and that in the case of solar system, spontaneous fluctuations 
are more efficient than environmental fluctuations to ensure the 
quantum-classical correspondence.  Finally, Section V summarizes the main 
results of this letter and discusses their possible consequences.

\section{Time evolution of the Wigner quasiprobability and logarithmic 
time}
	
Let us briefly review the arguments used by Zurek and Paz 
\cite{ZP94,ZP96,Z98} to derive the logarithmic time, and those used by them 
to settle the ensuing paradox as well.  It is convenient to adopt the 
Wigner formalism and write the time evolution of the quantum system in 
terms of the corresponding quasi-probability $W(x,p)$.  Furthermore, let us 
consider for simplicity the one-dimensional motion of a body under the 
influence of the analytical potential $V(x)$ and of a coupling, through the 
coordinate $x$, with a thermal environment.  The resulting evolution equation 
is \cite{CL83}  
\begin{eqnarray}
&&\dot{W} =\left\{H,W\right\}_{PB}+\sum_{n \geq 
1}^{\infty}\frac{(-\hbar^{2})^{n}}{2^{2n}(2n+1)!}\left(\frac{\partial^{2n+1}V}
{\partial x^{2n+1}}\right) \frac{\partial^{2n+1}W}
{\partial p^{2n+1}} \nonumber \\
&&+2\gamma \frac{\partial}{\partial p}\left(p 
W\right)+D\frac{\partial ^{2}W}{\partial p ^{2}} \;,
\label{wig1}
\end{eqnarray}
where $\left\{H,W\right\}_{PB}$ denotes the classical Poisson brackets, 
the second term is the non-classical, or Moyal term, and the last two terms 
describe the effects of the environment, producing (respectively) 
dissipation and diffusion.  The analysis of Zurek and Paz concerns the 
so-called ``reversible classical limit'', in which the relaxation 
rate $\gamma$  
tends to zero, with the diffusion coefficient 
$D= 2M \gamma KT $ kept constant.  Furthermore, they set $t \ll 1/\gamma $, 
so as to neglect the relaxation term of Eq.~(\ref{wig1}).  

Zurek and Paz set for the system an initial condition compatible 
with the assumption that this is a classical ``macroscopic'' state.  This 
means a Gaussian wave packet that in the corresponding classical phase 
space happens to be round and smooth over scales much larger than $\hbar$.  
Notice that the momentum and the coordinate widths, denoted by 
$\sigma _{p}(0)$ and 
$\sigma _{x}(0)$, respectively, are such that $\sigma _{x}(0) \sigma _{p}(0)
\gg  \hbar$.  Even if we set $D = 0$, the initial time evolution of the system is indistinguishable from that 
predicted by classical physics.  This is so because the second term on the 
r.h.s. of Eq.~(\ref{wig1}), the Moyal term, involves derivatives of a smooth 
function and $\hbar $ is negligible compared to the classical action, thereby 
resulting in extremely small corrections to classical physics.  However, as 
an effect of the fragmentation process resulting from the chaotic nature of 
the classical dynamics \cite{Z85}, the Moyal term becomes increasingly 
larger upon increase of time.  In fact classical evolution implies that the 
exponential contraction along a given phase-space direction is balanced by 
an exponential dilatation along another directions so as to fulfill the 
classical prescription that the phase-space volume has to be conserved.  
Let us assume, for instance, the momentum squeezing $\Delta p(t) = 
\sigma _{p}(0)\exp(-\lambda t)$, 
where $\lambda $ is the relevant Lyapunov exponent.  As a consequence, the 
quasi-probability $W(x,p)$ dilates in the $x$ direction so as to extend the 
quantum coherence to the distance $l(t) \simeq \hbar/\Delta p(t)$.  
When this coherence 
length $l(t)$ becomes of the order of the nonlinearity parameter 
$\chi $ defined by 
\begin{equation}
	\chi \simeq \sqrt{\frac{\partial V/\partial x}{\partial^{3} V/\partial 
	x^{3}}}   \;,
	\label{chi3}
\end{equation}
the Moyal term is no more negligible and quantum dynamics begin 
departing from classical physics.  Zurek and Paz show that this condition 
is realized at the logarithmic time $t_{Q}$, whose explicit expression is 
\begin{equation}
	t_{Q} = \frac{1}{\lambda}\ln\left(\frac{\chi 
	\sigma_{p}(0)}{\hbar}\right)
	\label{tq4} \;.
\end{equation}

As pointed out in Section I, this is a state of affair dangerous for 
the quantum-classical correspondence.  Thus, Zurek and Paz set $D > 0$.  
Under this new condition, the classical fragmentation cannot proceed 
indefinitely because the squeezing due to chaotic instability and the 
spreading due to diffusion will lead to a standoff characterized by 
\begin{equation}
	\sigma_{p}(0) \exp(-\lambda t) \simeq \sqrt{2Dt} \;;
	\label{stand5}
\end{equation}
this allows to estimate the time at which 
the fragmentation process is stopped, i.e.  
\begin{equation}
	t_{CG} = \frac{1}{\lambda}\ln\left(\frac{ 
	\sigma_{p}(0) \sqrt{\lambda}}{\sqrt{2D}}\right)
	\label{tcg6} \;.
\end{equation}
The condition necessary to 
make the quantum-classical correspondence valid is therefore 
\begin{equation}
	t_{CG} < t_{Q} \;.
	\label{cond7}
\end{equation}
If the 
noise intensity $D$ is large enough as to fulfill this condition, the Wigner 
quasiprobability never becomes so finely structured in momentum as to make 
the Moyal terms important.  The initial classical state remains smooth and 
so the quantum coherence length remains small.  

These arguments have been 
criticized by some authors \cite{CC95,FBA96}.  Casati and Chirikov 
\cite{CC95} noted that the original arguments of Zurek and Paz are 
oversimplified since they refer to an autonomous one-dimensional system, 
which cannot be chaotic.  They use an inverted parabola which is 
classically unstable but {\em nonchaotic} \cite{ZP94}.  A more serious criticism 
of \cite{CC95} is that the logarithmic time of Eq.~(\ref{tq4}) was already 
predicted by Berman and Zaslavski \cite{BZ78} and Berry {\em et al}.  
\cite{BBTV78} and that no significant departure of quantum from classical 
mechanics has been noticed to occur at this time: consequently no problem 
to the correspondence principle is created.  Casati and Chirikov point out 
the benefits of adopting the Husimi rather than the Wigner 
quasi-probability distribution.  The Husimi distribution is a Gaussian 
smoothing of the Wigner quasi-probability and the quantum fine structures 
are washed out by this coarse-graining procedure.  As a consequence, even 
if at a logarithmic time, according to Zurek and Paz, the Wigner 
quasi-probability significantly departs from the corresponding classical 
probability distribution, the Husimi distribution is still very close to 
the classical prediction, and consequently the correspondence principle is 
safe.  According to Casati and Chirikov, the departure of the classical 
from quantum predictions is expected to take place at the much longer 
relaxation time $t_{R}$, which is proportional to some power of $(A/\hbar)$ 
\cite{CC95}.  

It has to be pointed out, however, that the Husimi distribution 
is equivalent to the Wigner distribution, and that both are equivalent to 
ordinary quantum mechanics.  Thus, the validity of the correspondence 
principle cannot depend only on the adoption of the former rather than the 
latter distribution.  The reasons for the lack of any significant departure 
of quantum from classical predictions has been discussed in detail by 
Roncaglia {\em et al}.  \cite{RBWG95} and by Bonci {\em et al}.  
\cite{BGLR96} and the 
conclusion was reached that this is essentially due to the fact that a 
Gibbs ensemble picture is adopted and the attention is focused only on 
expectation values of dynamical variables, which are coarse-grained 
quantities, typically not very sensitive to quantum coherence effects 
\cite{notehz}.  

The work of Farini, Boccaletti and Arecchi \cite{FBA96} 
criticizes another aspect of the Zurek and Paz argument: classical chaos is 
characterized by stretching and folding and Ref. \cite{ZP94} neglects 
folding, i.e.  neglects the possibility of rapid twisting of the phase 
space direction corresponding to the local maximum Lyapunov exponent.  In 
this case a given variable, say the momentum, does not experience only 
squeezing but both expansion and contraction, depending upon the local 
position on the trajectory.  These authors supported their arguments by 
numerically studying a one-dimensional chaotic system, i.e.  a double-well 
potential under the influence of a coherent perturbation.  They found a 
parameter region in which the twisting process is important so that the 
non-classical Moyal term does not become comparable to the classical term 
at the transition time $t_{Q}$, or, equivalently, no transition from classical 
to quantum mechanics occurs at the time predicted by \cite{BZ78} and 
\cite{BBTV78} as well as by \cite{ZP94}.

\section{Mechanics and thermodynamics: a general perspective} 
	
In this Section we address the same problem as that of Section II from a 
different perspective, still heuristic, but more general than that adopted 
by Zurek and Paz.  This will serve the useful purpose of bypassing, at 
least in part, the criticisms of Refs.\cite{CC95,FBA96}.  In fact we shall 
show that the conclusions of Zurek and Paz are correct provided that the 
strong chaos condition is ensured.  

We adopt the same picture as that used 
by Zaslavsky in \cite{Z85} to illustrate the difference between purely 
ergodic motion and motion with mixing.  We consider an $N$-dimensional 
classically chaotic system and a single trajectory moving within the phase 
space region $S(E)$ compatible with a given energy $E$.  Let us denote 
by $M(t)$ 
the {\em unexplored area}, namely the area of the phase space region $S(E)$ 
not yet explored by this trajectory at a given time $t$.  In the mixing case 
\cite{Z85} it is expected that $M(t)$ tends to zero exponentially 
for $t$ tending to infinity.  

We think that a more general picture 
can be obtained 
by using the arguments by Tsallis and co-workers \cite{TPZ97,LT98}.  These 
authors are trying to establish a unified approach to nonlinear dynamics, 
extending from strong to weak chaos, based on the generalization of the 
property of exponential sensitivity to initial conditions.  Their arguments 
rest on the connection between this generalized property and the entropic 
index $q$ as well as on numerical calculations which support their 
conjecture.  By adapting their approach to the calculation of $M(t)$, we 
obtain \cite{noteonTSALLIS}.  
\begin{equation}
	M(t)=\frac{M(0)}{\left[1+\lambda_{q}t(1-q)\right]^{\frac{1}{1-q}}}
	\label{tsa8} \;.
\end{equation}
The case of strong chaos corresponds to 
$q = 1$, which makes the decay of $M(t)$ exponential, while the more general 
case of weak chaos corresponds to $q < 1$ \cite{noteonTSALLIS}.  If we now 
quantize this system, the quantum manifestations of the system can be 
neglected if the size of the wavepacket is small compared to $M(0)$.  
However, even in this case, it is not possible to disregard the quantum 
effects when $ M(t)$ becomes of the order of $(\hbar)^{N}$: 
at this time quantum 
interference between two different portions of the classical trajectory 
becomes important.  This condition defines the quantum-classical 
correspondence breakdown time $t_{Q}$ as: 
\begin{equation}
	M(t_{Q}) = \hbar^{N}
	\label{hba9} \;.
\end{equation}
It is straightforward to show that: 
\begin{equation}
	t_{Q} = \frac{1}{\lambda_{1}}\ln\left(\frac{M(0)}{\hbar^{N}}\right)
	\label{tq10} 
\end{equation}
if the strong chaos condition $q = 1$ is realized, and 
\begin{equation}
	t_{Q} = \frac{1}{\lambda_{q}(1-q)}\left[\left(\frac{M(0)}{\hbar^{N}}
	\right)^{1-q}-1\right]
	\label{tq11} 
\end{equation}
if the weak chaos condition $q < 1$ is fulfilled.  Therefore, we see from 
Eq.~(\ref{tq10}) that we recover the logarithmic dependence of 
\cite{ZP94,BZ78,BBTV78} 
on $\hbar$
(in the more general $N$-~dimensional case) only in the case of strong chaos, 
$q=1$, while in a more generic condition of weak chaos, $q<1$ , the dependence 
of $t_{Q}$ on $\hbar$ is given by an inverse power law, as shown by 
Eq.~(\ref{tq11}).  
In the case where the action $M(0)$ is macroscopic, this means a significant 
increase of the time of transition from classical to quantum mechanics.  

In our opinion, the above results endorse the arguments by Zurek and Paz and 
lend them protection from the criticism of Farini {\em et al}. \cite{FBA96} if 
not completely from those of Casati and Chirikov \cite{CC95}.  In the 
one-dimensional case $N=1$, our  prediction for $t_{Q}$ in the strong chaos case 
becomes practically identical to the Zurek and Paz result, Eq.~(\ref{tq4}).  
As to 
the criticism raised in \cite{FBA96}, Eqs.~(\ref{tq10}) and 
(\ref{tq11}) naturally suggest 
that the parameter region considered there, where the twisting process is 
important, actually might correspond to a situation of weak chaos in which 
the transition from classical to quantum physics is significantly postponed 
in time.  

Our argument does not make the conclusions of Zurek and Paz on 
the physical significance of $t_{Q}$ safe from the main criticism of Ref.  
\cite{CC95}, concerning, as we have seen, the fact that no significant 
difference between quantum and classical predictions shows up at the 
logarithmic time.  As explained in detail in \cite{RBWG95,BGLR96}, this 
criticism is essentially correct (see however \cite{notehz}) if applied to 
the measurement of smooth variables or to the observation of other 
coarse-grained quantities.  However, we cannot rule out the possibility 
that the breakdown of the correspondence principle, if no use of $D 
\neq 0$ is 
made, can be detected, in principle, by means of a different kind of 
observation.  Thus, in spite of the criticism of Casati and Chirikov, we 
believe that this breakdown would be signalled by the substantial 
difference between the Wigner function and the corresponding classical 
probability distribution.  We also note that this departure of the quantum 
from the classical distribution occurs at a relatively short time, if the 
strong chaos condition making $t_{Q}$ logarithmically dependent on the Planck 
constant is adopted.  

We think in fact that the use of the Husimi 
distribution, suggested in \cite{CC95}, is misleading.  It is well known 
that, because of its coarse-grained nature, the Husimi function (or 
$Q$ function in the quantum optics language) cannot be used as a convenient 
visual representation of the quantum coherence.  It fails signalling 
quantum coherence even when this becomes macroscopic.  The birth of quantum 
coherence, on the contrary, is clearly signalled by the wild oscillations 
of the Wigner function, and by its negative values as well.  A clear 
example is provided by the linear superposition of two spatially separated 
coherent states: while the Wigner function associated to it is quite 
different from that associated to the corresponding statistical mixture, 
with the use of the Husimi function we would be practically unable to 
distinguish one case from the other.  

More generally speaking, in our 
opinion, any formal mathematical procedure, inspired to the coarse-graining 
procedures, but not corresponding to a real physical process, independent 
of the observer, is unsatisfactory because it is ultimately justified only 
by the limitation of a given observation process, and the convolution used 
to process coarse graining can always be inverted so as to recover the 
original fine-grain picture.  Furthermore, even setting aside the 
possibility of overcoming the limitations of a given observation process, 
we are made to feel uncomfortable by a theoretical perspective where the 
correspondence principle is made safe by a human dependent coarse-graining 
process rather than by a genuine property of nature.  For these reasons we 
are forced to look for a theoretical perspective where the coarse graining 
process is made by nature, rather than by the observer.  

The need for this 
choice is reinforced if we move to treat the second of the two problems 
under discussion in this letter: the derivation of the second law of 
thermodynamics from mechanics.  As well known, the standard mechanical 
entropy, the Gibbs entropy associated to the Liouville density 
$\rho ({\bf X},t)$ of the classical phase space 
\begin{equation}
	S_{G}(t)=\int_{\Gamma}d{\bf X} \rho({\bf X};t)\ln \rho({\bf X};t)
	\label{gibbs12}  \;,
\end{equation}
is a constant of motion because of the 
conservation of phase space volumes in Hamiltonian systems.  
Irreversibility and, correspondingly, an increasing entropy (i.e., the 
second law of thermodynamics) can be derived from a dynamical picture only 
by replacing the Liouville density by its coarse-grained version 
$p ({\bf X},t)$, and this latter is obtained by averaging 
$\rho ({\bf X},t)$ over a suitably small 
phase-space volume.  In the case of a classically chaotic system this 
coarse-graining is a quite natural request for a human observer, since, if 
the mixing assumption is made \cite{Z85}, an initial condition given by a 
smooth Liouville density is changed into a highly fragmented distribution 
$\rho ({\bf X},t)$.  The coarse-graining process changes the Liouville density 
$\rho ({\bf X},t)$ 
into the smooth $p({\bf X},t)$, and this is indistinguishable 
from the conventional 
microcanonical distribution: as a remarkable effect of all this, the Gibbs 
entropy is naturally changed into the Boltzmann entropy.  

In our opinion, 
deriving thermodynamics from classical mechanics by making recourse to 
observer's limitations is unsatisfactory.  A much better perspective would 
be given by the coarse-graining process occurring as a genuinely phenomenon 
of nature rather than as a merely mathematical procedure reflecting 
observer's limitations.  The theoretical proposal of Zurek and Paz is 
attractive because it affords an illuminating example of physical (rather 
than subjective and merely mathematical) coarse-graining process.  In fact, 
the time $t_{CG}$ of Eq.~(\ref{tcg6}), 
at which the fragmentation process induced by 
chaos is stopped by diffusion, is not only the time that must become 
shorter than $t_{Q}$ to prevent the breakdown of the quantum-classical 
correspondence.  This is also the time at which the coarse-graining 
process, produced by nature, changes the incompressible Hamiltonian fluid 
into a fluid dynamic breaking the Liouville theorem, thereby making the 
entropy of the chaotic system increase \cite{ZP94,Z98}.  This is the reason 
why we have used the notation $t_{CG}$.  This time signals the transition from 
mechanics to thermodynamics.  

This coarse-graining time can be derived in a 
general way using again the above argument based on the unexplored area 
$M(t)$.  In fact, in the presence of an isotropic diffusion process with 
diffusion coefficients $D_{x}$ and $D_{p}$ for the space and the momentum, 
respectively, the phase space is practically filled by the trajectory and a 
coarse-graining process is completed when the condition 
\begin{equation}
	M(t_{CG})=\left(2\sqrt{D_{x}D_{p}} t_{CG}\right)^{N}
	\label{em13}
\end{equation}
is realized.  
Solving the equation for $t_{CG}$, we obtain 
\begin{equation}
	t_{CG} = \frac{1}{\lambda_{1}}\ln\left(\frac{M(0)\lambda_{1}^{N}}
	{\left(2\sqrt{D_{x}D_{p}}\right)^{N}}\right)
	\label{tcg14} 
\end{equation}
if the case of strong chaos, and 
\begin{equation}
	t_{CG} = \frac{1}{\lambda_{q}(1-q)}\left[\left(\frac{M(0)\lambda_{q}^{N}}
	{\left(2\sqrt{D_{x}D_{p}}\right)^{N}}
	\right)^{1-q}-1\right]
	\label{tcg15} 
\end{equation}
in the case of weak chaos.  We see that, as in the case of $t_{Q}$, the 
adoption of the weak chaos condition makes $t_{CG}$ much larger (notice that 
$M(0)$ is a macroscopic quantity) and that its expression reduces to that of 
Eq.~(\ref{tcg6}), the same as that derived by Zurek and Paz 
in the one-dimensional 
case, if the assumption of strong chaos is made and only the momentum 
variable is assumed to undergo the influence of external fluctuations.

\section{Spontaneous fluctuations as a source of the 
classicality of the solar system dynamics}

In the preceding Section we have seen that the introduction of fluctuations 
(of environmental origin according to the theoretical perspective adopted 
in Refs.  \cite{ZP94,ZP96,Z98}) is equivalent to establishing a 
coarse-graining process, and that this yields a natural derivation of the 
second law of thermodynamics from mechanics.  In this section we prove that 
the same important effects are produced by fluctuations of different 
origin: these are those postulated by the so-called spontaneous reduction 
models \cite{GRW,GPR,PD,GGR}.  These models are modifications of quantum 
mechanics with negligible effects at the microscopic level, where standard 
quantum theory still holds true.  At the macroscopic level, on the 
contrary, these corrections cause a fast collapse of the linear 
superposition of distinct physical properties, so that, in accordance with 
our daily experience, any single state is associated to only one value of a 
given physical property: the paradoxical condition denoted as Schr\"odinger's 
cat is made impossible.  The first model of this kind was proposed by 
Ghirardi, Rimini e Weber \cite{GRW}, who assumed that each particle is 
subject to abrupt and spontaneous localizations with mean rate 
$\lambda_{GRW} \simeq 10^{-16}$ sec$^{-1}$ and localization length 
$a \simeq 10^{-5}$ cm.  The rate $\lambda_{GRW}$ and the length 
$a$ are to be regarded as two new constants of nature.  We shall refer to 
this as GRW model.  This model was later improved by Ghirardi, Pearle and 
Rimini \cite{GPR} to take into account that the particles are 
indistinguishable, and the instantaneous localization process was replaced 
by a continuous version described by a Wiener process with intensity 
$\gamma_{GPR} \simeq 10^{-30}$ cm$^{-3}$ sec$^{-1}$.  
This model will be denoted by us as GPR model.  At the 
same time, Di\'osi and independently Penrose, \cite{PD}, tried to elaborate a 
reduction model based on gravity without introducing any new constant of 
nature; however Ghirardi, Grassi and Rimini \cite{GGR} showed that these 
models resulted in an exceedingly large energy increase, and amended the 
model from this fault by introducing a mass smearing over volumes of the 
order of $a^{3} \simeq 10^{-15}$ cm$^{3}$ ($a$ is again the new constant 
that according to the 
GRW theory establishes the space scale within which the wave function 
reduction occurs), so as to keep the energy increase within experimental 
constraints.  Therefore, this latter version of the model of Ref.  
\cite{GGR} only requires one new constant of nature, the length $a$, is 
perfectly consistent with experimental constraints and can be physically 
explained as an effect of spontaneous gravitational potential 
fluctuations of dipole type \cite{PS95}.  We shall refer to this version of 
the theory of spontaneous fluctuations as GGR model.  

The introduction of a 
new stochastic term in the Schr\"odinger equation is equivalent to adding a 
new, and fundamental, diffusion-like process to the dynamics of systems and 
for this reason the effects of spontaneous reduction models are analogous 
to those produced by the environment-induced decoherence.  The main 
difference with standard thermal master equations (of environmental origin) 
is that no friction term is associated to these reduction models, and this 
is responsible for the steady energy increase, a property characterizing 
all these models.  To illustrate formally the nature of these models, let 
us consider the quantum dynamics of the center of mass of a macroscopic 
body of mass $M$, volume $V$ and with a number $N$ of constituent particles.  
According to all the reduction models the dynamics of the center of mass is 
almost completely decoupled from the internal dynamics.  Thus we can 
represent the dynamics of the center of mass coordinate $\vec{q}$ by means of an 
equation of motion for its reduced density matrix 
$\rho(\vec{q},\vec{q}\,',t)$.  This equation 
reads: , 
\begin{equation}
	\frac{\partial \rho(\vec{q},\vec{q}\,',t)}{\partial t}=
	\frac{i}{\hbar}\left\langle 
	\vec{q}|\left[\rho,H_{CM}\right]|\vec{q}\,'\right\rangle - 
	\Gamma\left(|\vec{q}-\vec{q}\,'|^{2}\right)\rho(\vec{q},\vec{q}\,',t)
	\label{roq16} \;,
\end{equation}
where $\Gamma $ is a function of $|\vec{q}-\vec{q}\,'|^{2}$ 
dependent upon the chosen 
reduction model.  The explicit form of $\Gamma $ is: 
\begin{equation}
	\Gamma_{GRW}\left(|\vec{q}-\vec{q}\,'|^{2}\right)=
	N\lambda_{GRW}\left(1-\exp\left[-\frac{|\vec{q}-\vec{q}\,'|^{2}}{4a^{2}}
	\right]\right)
	\label{grw17}
\end{equation}
for the original GRW model \cite{GRW}, 
\begin{equation}
	\Gamma_{GPR}\left(|\vec{q}-\vec{q}\,'|^{2}\right)=
	\frac{\gamma_{GPR}}{2}
	\int d^{3}x \left[F(\vec{q}-\vec{x})-F(\vec{q}\,'-\vec{x}')\right]^{2}
	\label{gpr17}
\end{equation}
for the GPR model of Ref. \cite{GPR}, where $F(\vec{q})$ is a 
function depending on the particle number density at the point 
$\vec{q}+ \vec{y}$, $D(\vec{y})$ 
\cite{GPR}, defined by: 
\begin{equation}
	F(\vec{q}-\vec{x})=
	\int d^{3}y \frac{D(\vec{y})}{(2\pi)^{3/2} 
	a^{3}}\exp\left[-\frac{|\vec{q}-\vec{x}+\vec{y}|^{2}}{2a^{2}}\right]
	\label{gpr18} \;; 
\end{equation}
finally for the GGR model, namely the 
gravitationally-induced reduction model of \cite{PD} amended from the 
excessive energy increase in \cite{GGR}, we have 
\begin{equation}
	\Gamma_{GGR}\left(|\vec{q}-\vec{q}\,'|^{2}\right)=
	-\frac{1}{\hbar} \left[U(0)-U(|\vec{q}-\vec{q}\,'|)\right]
	\label{ggr19} ;\,
\end{equation}
where $U(|\vec{q}-\vec{q}\,'|)$ is 
the gravitational interaction energy between two copies of our macroscopic 
body with the center of mass placed at $\vec{q}$ and $\vec{q}\,'$.  
The evolution equation Eq.~(\ref{roq16}) 
clearly shows that the new term induces a rapid decay of the quantum 
coherence between wave-functions components separated by distances 
larger than the new length constant $a$.  

These correction terms are nothing 
but diffusion-like terms and this can be clearly seen if one rewrites 
Eq.~(\ref{roq16}) for $\rho(\vec{q},\vec{q}\,',t)$, 
in terms of the Wigner function 
$W(x,p)$.  In fact multiplication by $|\vec{q}-\vec{q}\,'|^{2}$ 
in the coordinate representation 
is equivalent to the application of the Laplacian operator in momentum 
space in the Wigner representation \cite{CL83},
\begin{equation}
	|\vec{q}-\vec{q}\,'|^{2}\rho(\vec{q},\vec{q}\,') \leftrightarrow 
	-\nabla ^{2}_{p} W(\vec{x},\vec{p})
	\label{nabl20} ;\,
\end{equation}
and therefore these 
spontaneous reduction models are equivalent to adding the diffusion-like 
operator 
\begin{equation}
	-\Gamma\left(-\nabla ^{2}_{p}\right) W(\vec{x},\vec{p})
	\label{nabl21} 
\end{equation}
to the evolution equation of the Wigner function ($\Gamma $ can 
always be expressed as a power series and $\Gamma(0)=0$).  
Since the effects of 
environmental decoherence are established by Zurek and Paz in the 
reversible classical limit, where dissipation is negligible compared to 
diffusion, we can apply to the new cases of this Section the same arguments 
as those illustrated in Section II, with no problem whatsoever.  This is 
equivalent to stating that also the spontaneous fluctuations allow us to 
protect the classical-like system from an early transition to the quantum 
regime, induced by deterministic classical chaos.  It is possible to 
realize the key condition $t_{CG} < t_{Q}$, which makes it impossible for the 
transition from classical to quantum mechanics to occur at any time, so as 
to make the quantum-classical correspondence safe.  In fact for both the 
spontaneous reduction models and the environment-induced decoherence we 
derive the following evolution equation for the dispersion $\Delta 
p_{i}(t)$ of a 
given momentum component of the macroscopic body , 
\begin{equation}
	\frac{d}{d t} \left(\Delta p_{i}(t)\right)^{2}=
	\left\langle \ldots \right\rangle _{Q}+2D
	\label{dif22} \;,
\end{equation}
where $\langle \ldots \rangle _{Q}$ stands 
for the quantum Hamiltonian terms and the diffusion coefficient $D$ depends 
on the model chosen according to the prescriptions of Table \ref{table1}.
\begin{table}[t]
\caption{Expression of the diffusion coefficients
for the various models. The subscripts env, GRW, GPR and GGR denote the 
fluctuation intensities predicted by refs.\protect\cite{ZP94,GRW,GPR,GGR}, 
respectively.  $G$ is the gravitational constant and $S_{\protect\perp}$
denotes the surface of the body, transverse to the direction of $p_{i}$.}
\label{table1}
\begin{center}
\begin{tabular}{| l | c | c | r |}
$D_{env}$ & $D_{GRW}$ & $D_{GPR}$ & $D_{GGR}$ \\
\hline
$2M\gamma kT$ & $\frac{N\lambda_{GRW}\hbar^{2}}{4a^{2}}$ & 
$\frac{\gamma_{GPR}\hbar^{2}N^{2}S_{\perp}}{4V^{2}a\sqrt{\pi}}$ & $\frac{G\hbar 
M^{2}}{2V}$ \\ 
\end{tabular}
\end{center}
\end{table}
The resulting linear increase in time of the momentum dispersion due to 
diffusion is the key ingredient for the determination of $t_{CG}$.  Using the 
same arguments as those of Sections II and III we derive for $t_{CG}$ the same 
expression as that of Eq.~(\ref{tcg6}), with the same dependence on $D$.  The 
prescriptions to adopt to evaluate $D$ is dictated by the model adopted, as 
indicated in Table \ref{table1}.  

Let us now consider again the classically chaotic 
celestial systems described in the introduction and discussed in 
\cite{Z98}, and let us focus on the solar system.  According to the 
discussion of the preceding Section, if the motion of the solar system as a 
whole satisfies the strong chaos condition (with the global Lyapunov 
exponent evaluated in \cite{L89,SW92}), the logarithmic time prediction 
derived by Zurek in \cite{Z98} is correct and the present quantum state of 
the solar system would be characterized by planetary wave functions 
spreading all over their orbits.  However, as shown above, even a tiny 
diffusion process is sufficient to recover classicality, namely a time 
$t_{CG}$ 
shorter than the breakdown time $t_{Q}$.  When making the estimates discussed in 
\cite{Z98}, Zurek considers for simplicity only the motion of the most 
massive planet, Jupiter, and considers the environmental diffusion due to 
the interaction of Jupiter with the interplanetary gas in its vicinity (of 
approximate number density $0.1$ cm$^{-3}$).  This yields $\gamma \simeq 
10^{-26}$ sec$^{-1}$. Then, 
using Jupiter surface temperature $T \simeq 100\; ^{o}$K, we get 
\begin{equation}
D_{env} \simeq 10^{-10} \;{\rm erg \cdot gr/sec} 
\label{denv23} \;.
\end{equation}
By identifying the initial momentum dispersion $\sigma_{p}(0)$ 
with Jupiter's 
mean momentum, and so overestimating $t_{CG}$, we get from 
Eq.~(\ref{tcg6}) $t_{CG} 
\simeq 3\cdot 10^{8}\; {\rm 
years} \; <  t_{Q} \simeq 7\cdot 10^{8}$ years.  
Although the coarse-graining time is extremely 
long, the transition to quantum mechanics is made impossible by the 
condition $t_{CG} < t_{Q}$: initial classical dynamics of the solar system will 
always remain classical.  Now we can apply the same estimate procedure to 
the diffusion induced by spontaneous fluctuations, using the expressions 
for the corresponding diffusion coefficients given by Table \ref{table1}. Using again 
Jupiter values for $M$, $V$ and $N$, and the spontaneous reduction parameter 
values, we obtain the values of Table \ref{table2}.
\begin{table}[b]
\caption{Numerical estimates for the diffusion coefficients for the 
environmental decoherence and for the
spontaneous reduction models expressed in erg$\protect\cdot$gr/sec. 
The meaning of the subscripts is the same as 
that of the corresponding subscripts in Table \protect\ref{table1}.}
\label{table2}
\begin{center}
\begin{tabular}{| l | c | c | r |}
$D_{env}$ & $D_{GRW}$ & $D_{GPR}$ & $D_{GGR}$\\
\hline
$ 10^{-10}$ & $ 10^{-8}$ & $ 10^{-11}$ & $ 10^{-4}$ \\
\end{tabular}
\end{center}
\end{table}
With the joint use of the results reported in Table \ref{table1} 
and of the same 
argument as that applied to the case of environmental fluctuations, all 
these models are proved to share with the theory of environmental 
fluctuations the attractive property of yielding $t_{CG} < t_{Q}$, making it 
impossible for the solar system to fall in a quantum condition.  This 
is a natural consequence of the fact that, whatever the origin of 
fluctuation is, either spontaneous or environmental, the dynamics of 
the system is described by Eq.~(\ref{wig1}), with the friction term neglected.  
{\em The most interesting and surprising fact is that this celestial 
dynamics example is the first realistic situation in which spontaneous 
reduction effects (especially those of the gravitational model) are 
found to be much stronger than those associated to environment-induced 
decoherence}.  In fact a number of estimates of spontaneous reduction 
effects exist in literature \cite{BGM95,TEG91,TVG95,BJK98,R95} and all 
of them agree on the fact that these effects are blurred by 
environmental decoherence.  This fact has led to the common criticism 
that these models are a useless modification of quantum mechanics.  
This would be so because these theories are made to pass all the 
experimental tests by an {\em ad hoc} choice of parameters, which, at the 
same time, makes them practically unverifiable \cite{TEG91}, since 
their effect is always blurred by decoherence.  

However, we have to 
point out that the results concerning the solar  system
here discussed are not a compelling evidence of the gravitationally-induced 
objective reduction model, since no experiment can be made to assess if the 
solar system dynamics are made classical by the environment-induced 
decoherence or by the spontaneous gravitational fluctuations.  
Is there, in 
this condition, any deep reason to trace back the fluctuations necessary to 
maintain the classicality of macroscopic systems to the corrections to 
ordinary physics rather than to the environmental influence?  The answer to 
this important question is in positive.  Let us see why.  First of all, as 
explained in \cite{PT93,BGM95} the theory of environmental decoherence 
provides only a statistical solution to the problem of the emergence of 
classicality, in the sense that it only shows the emergence of statistical 
mixtures of classically distinguishable events from quantum dynamics.  This 
theory does not explain the objectification problem \cite{BLM91}, the 
uniqueness of events, i.e.  how a single localized event out of the many 
which are statistically possible is realized.  On the contrary, this 
emergence of individual events is well explained by spontaneous reduction 
models whose fundamental equation is a stochastic modification of the 
Schr\"odinger equation: the emergence of an event corresponds to a single 
stochastic process in the Hilbert space of the Universe.  Another argument 
making objective reduction theories preferable, in our opinion, to the 
approaches based on environmental decoherence is connected to the problem 
described in section III, i.e.  the derivation of the second law of 
thermodynamics from mechanics, as an effect of the coarse-graining on the 
classical phase space.  As we have already said, a process of coarse 
graining realized by nature is more satisfactory than an equivalent 
mathematical procedure forced by the observation limitations.  The theory 
of environment decoherence is not totally free from this subjective aspect.  
First of all, we note that the environmental fluctuation is derived from 
within the ordinary quantum mechanics by making a contraction on the 
environmental degrees of freedom: a subtle way of stating the limitations 
affecting human observation.  Moreover the diffusion and the relaxation 
terms are derived assuming the environment to be in a state of 
thermodynamical equilibrium.  This means that an implicit thermodynamical 
assumption is made, even if this serves only the purpose of realizing a 
coarse-graining condition.  We cannot conclude that thermodynamics is 
genuinely derived from dynamics, since the known procedures used to derive 
fluctuation-dissipation processes rest already on thermodynamics.  It is 
better to make a frank admission that the assumptions made correspond to 
correcting ordinary physics, and that the diffusion process necessary to 
realize the ``objective coarse-graining'' are something new, a stochastic 
seed foreign to ordinary dynamics.  This frank admission is made explicit 
by the theories of spontaneous fluctuations.

\section{Conclusions}   

The main results of this letter are the following: 
\begin{enumerate}
\item{We have presented a 
general, even though heuristic, argument, confirming the logarithmic 
dependence on $\hbar$ of the time of transition from classical to quantum 
mechanics, established by the authors of Refs.  \cite{ZP94,ZP96,Z98}. The 
derivation of this letter shows that the condition of strong chaos is a 
fundamental requirement for this prediction to be correct, and make the 
original conclusions of Zurek and Paz more robust against the criticism of 
\cite{CC95,FBA96}}.  
\item{We have shown that the gravitationally-induced 
objective reduction model of \cite{PD,GGR}, applied to the chaotic dynamics 
of the solar system, yields a transition from quantum to the classical 
domain, which is much faster than that produced by the environmental 
fluctuations.}
\end{enumerate}  

This fact is not a compelling evidence of the existence of 
spontaneous fluctuations.  In fact, the expression of the various diffusion 
coefficients (see Table \ref{table1}) shows that one needs very massive objects 
with an 
extremely weak dissipation (like planets) to make the objective reduction 
effects prevail on environmental decoherence: a condition certainly 
exceeding those of the experimental observation in an ordinary laboratory.  
Thus, it seems impossible at the moment, to assess experimentally whether 
the classical motion of macroscopic systems close to the conditions of the 
solar system, is due to environmental or to spontaneous decoherence.  We 
note furthermore that the condition of weak chaos has the effect of 
postponing both the transition from classical to quantum mechanics and the 
coarse graining time $t_{CG}$.  However, in this condition the transition time 
turns out to be much more sensitive to the intensity of the fluctuations.  
This means that, in principle, it would be much easier to distinguish the 
environment-induced decoherence process, whose intensity is expected to be 
temperature dependent, from the influence of spontaneous collapses, whose 
intensity is temperature independent.  In other words, weak chaos might 
offer the proper conditions to assess experimentally the existence of 
spontaneous fluctuations.


\begin{thebibliography}{99}
\bibitem{noteonTSALLIS}We make the plausible assumption that 
$M(t)$ is inversely proportional to the increase of the distance between two 
distinct trajectories starting from very close initial conditions.  Tsallis 
and co-workers \cite{TPZ97,LT98} study also the dynamics corresponding to 
attractive fixed points, with $q > 1$.  For the main purpose of this letter 
only the case $q < 1$ is of interest.
\end{thebibliography}
\end{document}